\documentclass[%
 aip,
 amsmath,amssymb,
 reprint,
 floatfix
]{revtex4-1}

\usepackage{graphicx}
\usepackage{dcolumn}
\usepackage{bm}

\usepackage[T1]{fontenc}
\usepackage{mathptmx}
\usepackage{booktabs}

\begin{document}

\title{Design of a High-Resolution Rayleigh-Taylor Experiment with the Crystal Backlighter Imager on the National Ignition Facility}

\author{A. M. Angulo}
 \affiliation{University of Michigan, Ann Arbor, Michigan 48109, USA}

\author{S. R. Nagel}%
\affiliation{Lawrence Livermore National Laboratory, Livermore, California 94550, USA}

\author{C. M. Huntington}%
\affiliation{Lawrence Livermore National Laboratory, Livermore, California 94550, USA}

\author{C. Weber}%
\affiliation{Lawrence Livermore National Laboratory, Livermore, California 94550, USA}

\author{H. F. Robey}%
\affiliation{Los Alamos National Laboratory, Los Alamos, New Mexico, 87545, USA}

\author{G. N. Hall}%
\affiliation{Lawrence Livermore National Laboratory, Livermore, California 94550, USA}

\author{L. Pickworth}%
\affiliation{MAX IV Laboratory, Lund University, P.O. Box 118, SE-22100 Lund, Sweden}

\author{C. C. Kuranz}
\affiliation{University of Michigan, Ann Arbor, Michigan 48109, USA}

\begin{abstract}

The Rayleigh-Taylor (RT) instability affects a vast range of High Energy Density (HED) length scales, spanning from supernova explosions (10$^{13}$ m) to inertial confinement fusion (10$^{-6}$ m).  In inertial confinement fusion, the RT instability is known to induce mixing or turbulent transition, which in turn cools the hot spot and hinders ignition.  The fine-scale features of the RT instability, which are difficult to image in HED physics, may help determine if the system is mixing or is transitioning to turbulence.  Earlier diagnostics lacked the spatial and temporal resolution necessary to diagnose the dynamics that occur along the RT structure. A recently developed diagnostic, the Crystal Backlighter Imager (CBI), \cite{Hall:2019, DoZonePlate} can now produce an x-ray radiograph capable of resolving the fine-scale features expected in these RT unstable systems. This paper describes an experimental design that adapts a well-characterized National Ignition Facility (NIF) platform to accommodate the CBI diagnostic. Simulations and synthetic radiographs highlight the resolution capabilities of the CBI in comparison to previous diagnostics. The improved resolution of the system can provide new observations to study the RT instability's involvement in mixing and the transition to turbulence in the HED regime.

\end{abstract}

\maketitle

\section{Introduction}

Late-time, nonlinear Rayleigh-Taylor (RT) unstable behavior plays a key role during multiple phases of inertial confinement fusion capsule implosions.  Imperfections present on the capsule interface cause mixing that cools the hot spot which reduces the fusion yield. \cite{Remington:2006,Robey:2003, RPD, Nagel:2017, Smalyuk:2017, Hurricane_2016, Clark_2019}  As the RT instability develops, the cold, dense material (spikes) interpenetrates the hot, less dense material (bubbles) and cools the system. However, due to diagnostic limitations it is difficult to diagnose the dynamics that occur along the spike tip and ascertain whether mixing is solely present, or if the system transitions to turbulence. Simulations are one method to observe this mixing behavior, but it is also difficult to model fine-scale structures. It is computationally expensive to  directly resolve the range of spatial and time scales necessary to diagnose mixing or turbulence. Reduced model simulations remove small-scale information and use numerical schemes to model small-scale effects on broader dynamics. Codes with identical initial conditions, but different numerical schemes generate different structures, \cite{darlington2001study, miles2004bubble} and therefore require experiments to validate these models.

\begin{figure}
 \begin{center}
 \includegraphics[width = 3.2in]{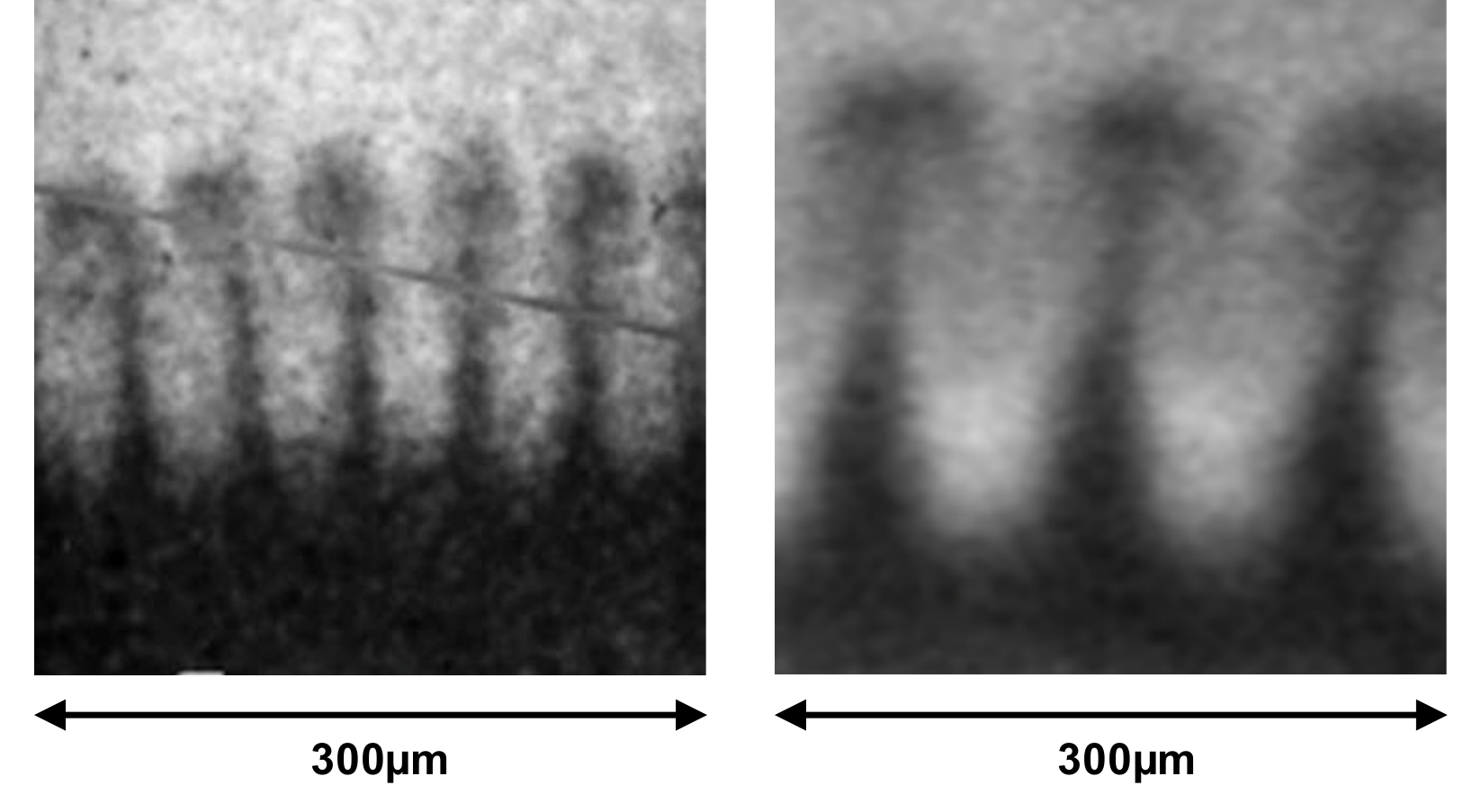}
 \caption{X-ray radiographs of RT instability structure obtained by Robey et. al on the Omega Laser in 2001 (left) and Nagel et. al on the NIF in 2016 (right). The larger initial wavelength (right) produces larger, more visible vortex features. } 
 \label{fig:Reshock Data}
 \end{center}
 \end{figure}

 Numerous experiments have been conducted at the Laboratory for Laser Energetics and the National Ignition Facility (NIF) that were aimed at measuring the nonlinear RT growth using x-ray radiography (Figure \ref{fig:Reshock Data}). \cite{Nagel:2017, Robey:2003, Zhou:2017}  Yet these platforms were optimized at measuring the gross mix width of the instability, and not the fine-scale features that develop along the spike tip.  Facility and diagnostic advancements now allow us to reach regimes and image features that may play a significant role in HED RT- unstable systems.

The remainder of this paper is organized as follows. In Section 2, we summarize the RT instability, vorticity, and the transition to turbulence in an RT-unstable HED system.  We provide a definition for turbulence as it relates to this experiment and discuss the Liepmann-Taylor scales for HED-laboratory-scale experiments in general. In Section 3, we provide an experimental design, referred to as the Vortex experiment,  that adapts the drive and diagnostic configuration of a related NIF campaign. In Section 4, we describe the CBI, a recently-commissioned x-ray radiography diagnostic that greatly improves spatial resolution on the NIF. Finally, in Section 5, we show a series of synthetic radiographs that are post-processed to emulate data produced by the CBI as it compares to a pinhole imager.  These synthetic radiographs inform any necessary improvement to achieve our desired resolution.

\section{Turbulence induced by the Rayleigh-Taylor Instability }

A hydrodynamic system is considered RT unstable when a density gradient opposes a pressure gradient ($\nabla \rho \cdot \nabla P < 0$).  In a system where a heavy fluid is accelerated into a light fluid, the contact surface is unstable and small perturbations initially present on the surface begin to grow. \cite{Rayleigh, taylor1950instability} Imposing a single-mode sine function of known wavelength and amplitude on the interface seeds the instability and enables a less-complicated description of how the penetrating heavy fluid (spikes) and light fluid (bubbles) evolve through various stages. \cite{jacobs1996experimental,roberts2016effects}  
 
  \begin{figure}
 \begin{center}
 \includegraphics[width = 3.2in]{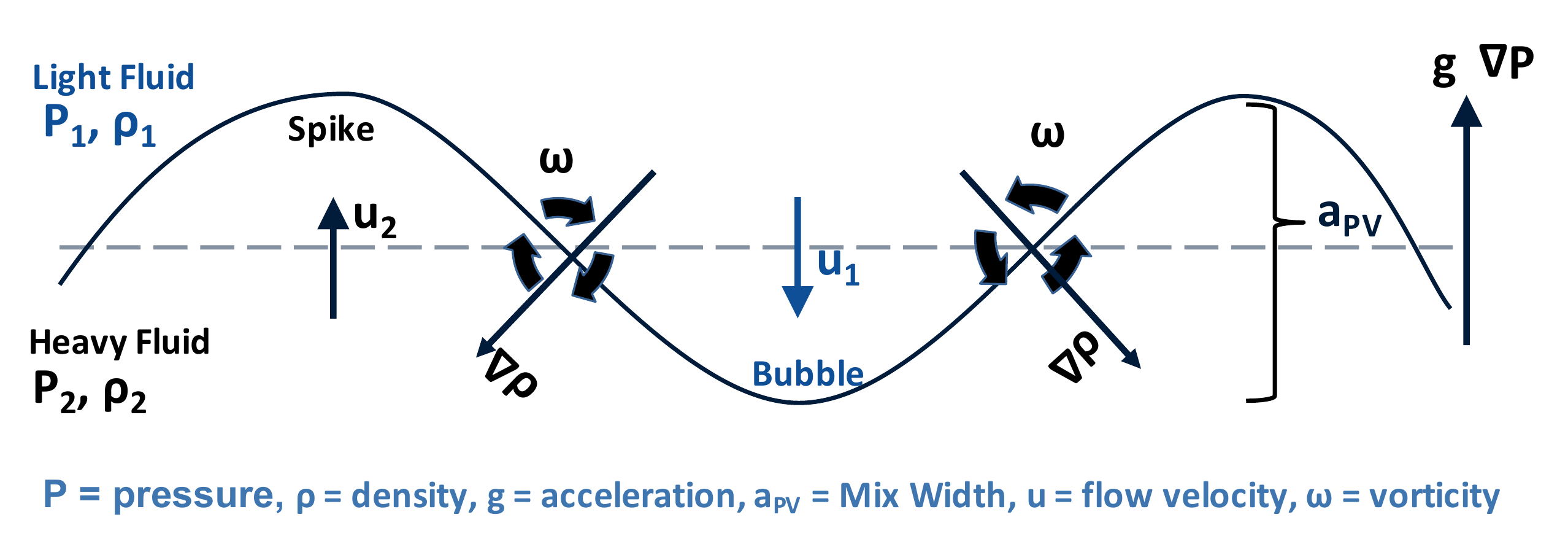}
 \caption{Diagram of the Rayleigh-Taylor Instability } 
 \label{fig:RTdiagram}
 \end{center}
 \end{figure}
 
As the instability further evolves and the amplitude grows to approximately 10 percent of the wavelength, it approaches the early nonlinear stage. \cite{Zhou:2017, zhou2019turbulent} It is during this stage that nonlinear terms become large enough to make significant contributions to the interface features that affect the growth rate.\cite{Zhou:2017, robey2003onset}  Vorticity, the curl of the velocity,  is introduced as the density gradient becomes misaligned from the pressure gradient along the length of the spike (Figure \Ref{fig:RTdiagram}).  It creates vortex features that cause the spike to broaden as it grows. It is also possible that the vorticity induces a transition to turbulence within this vortex feature, however this is not a guaranteed result.    \cite{jacobs1996experimental}  This paper describes an experimental design that aims to improve experimental resolution to begin to diagnose and capture turbulent transition within the vortex structure.

Turbulence has been studied quite prolifically throughout the century and has produced a multitude of definitions across a wide range of subfields of physics.  \cite{RPD, dimotakis2000mixing,  dimonte2006k}  Fundamental turbulence theory defines turbulence as  (i) the development of substantially convoluted structure at an interface, having spectral content or spatial extent far beyond those of the initial state, (ii) the appearance of an inertial range in the fluctuation spectrum, with power-law decay of the spectral energy density, and (iii) the development of strong mixing as indicated by supra-linear growth of the thickness of the mixing layer in time. \cite{RPD}


However, the experiment described in this paper highlights the practical aspect of understanding and diagnosing turbulence as it applies specifically to high-energy-density (HED) physics and to ICF. Therefore, we claim our system is transitioning, or has transitioned, to turbulence if we achieve a sufficiently high Reynolds number ($Re \sim 10^4$). \cite{RPD, dimotakis2000mixing, dimonte2006k,Robey:2003, Zhou:2017, zhou2019turbulent}  The Reynolds number is the ratio of inertial forces to viscous forces used to classify the type of flow. It is defined as

\begin{equation}
Re = \frac{uL}{\nu},
\label{eqn:ReynoldsNumber}
\end{equation}

where $u$ is the characteristic  velocity [$\mu$m/ns], $L$ is the characteristic length scale [$\mu$m], and $\nu$ is the fluid viscosity [$\mu$m$^2$/ns]. According to equation \ref{eqn:ReynoldsNumber}, if the Reynolds number is low, viscous effects are apparent, and the flow is considered laminar.   Conversely, if the Reynolds number is high, viscous effects are minimal or nonexistent, and the flow is turbulent.

 A method to experimentally ascertain that the system has reached $Re\sim 10^4$ is to resolve spatial scales corresponding to the Liepmann-Taylor scale,  $\lambda_{\rm{LT}} = 5LRe^{-1/2}$. \cite{zhou2019turbulent, zhou2003progress, zhou2007unification} The $\lambda_{\rm{LT}}$, or inertial range, is an intermediate length scale at which the fluid is decoupled from small-scale fluid viscosity and large-scale bulk dynamics. So, length scales which are larger than  $\lambda_{\rm{LT}}$ are not strongly affected by viscosity. Conversely, turbulent motions below  $\lambda_{\rm{LT}}$ are dominated by viscous forces and kinetic energy is dissipated into heat.  \cite{dimotakis2000mixing, Robey:2003} Additionally, the  appearance of $\lambda_{\rm{LT}}$ is one of the definitions of turbulence discussed earlier.

  \begin{figure}
 \begin{center}
 \includegraphics[width = 3.2in]{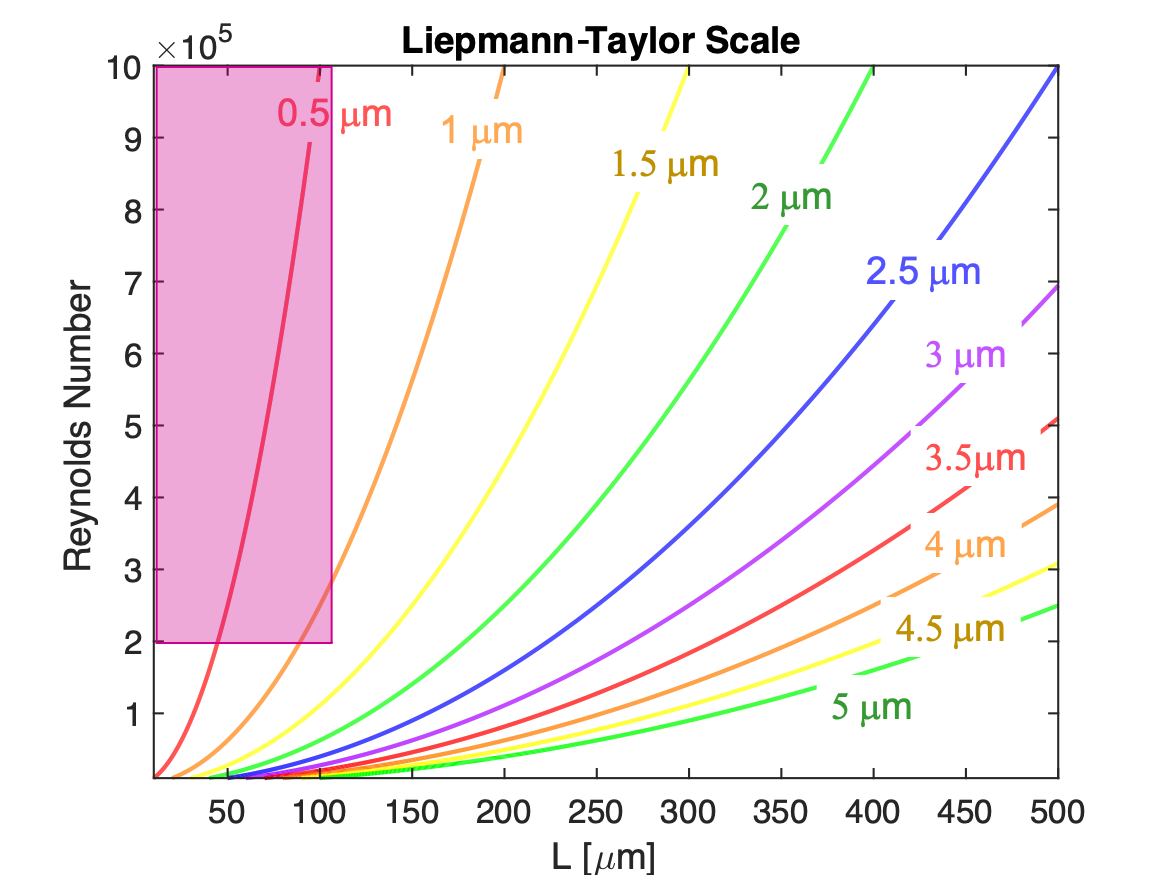}
 \caption{This figure shows Liepmann-Taylor scale as it relates to the Reynolds Number and HED-laboratory length scales (L). The region in parameter space where present experiments exist is highlighted by the red box. } 
 \label{fig:LTscale}
 \end{center}
 \end{figure}

 In the RT instability, length scales of order $\lambda_{\rm{LT}}$ would materialize within the diffusion layer that grows along the boundaries of the RT spikes and within the vortex structure.\cite{Robey:2003} For a HED-laboratory length scales (L $\sim$  50 - 100 $\mu$m, u $\sim$ 40 - 60  $\mu$m/ns and $\nu$ $\sim$ 10$^{-3}$ $\mu$m$^2$/ns), RT unstable experiments are estimated to produce a Reynolds number of order 10$^5$. A Reynolds number of order 10$^5$ yields a $\lambda_{\rm{LT}}$ of order 1 $\mu m$ -- one order of magnitude below present diagnostic capabilities. As shown in Figure \ref{fig:LTscale}, an experimental platform should allow $Re$ to remain as close to the 10$^4$ threshold as possible to maximize $\lambda_{\rm{LT}}$ within the HED parameter space.

\section{Experimental Design}
\noindent This experimental design utilizes a similar drive and physics package from previous successful campaigns conducted on the NIF. \cite{Nagel:2017, Robey:2003} This experiment is conducted in a rectangular shock tube and is driven by a halfraum on one end.  When the halfraum receives a laser pulse, it produces a thermal bath of soft x-rays that generates a strong shock into a physics package comprised of high density plastic and low density foam. The strong shock turns this initially solid target into a plasma and the material boundary between the plastic and foam becomes a fluid interface. The high density fluid, previously the plastic, is accelerated into the low density fluid, previously the foam, and the system becomes RT unstable.  A precise single-mode sine pattern is machined at the fluid interface to seed the RT instability with a known amplitude and wavelength such that it will achieve a sufficient degree of nonlinearity during the experiment.   After a specified time delay, additional lasers will irradiate a foil to produce the hard x-rays necessary to radiograph the evolving interface.

  \begin{figure}
 \begin{center}
 \includegraphics[width = 3. in]{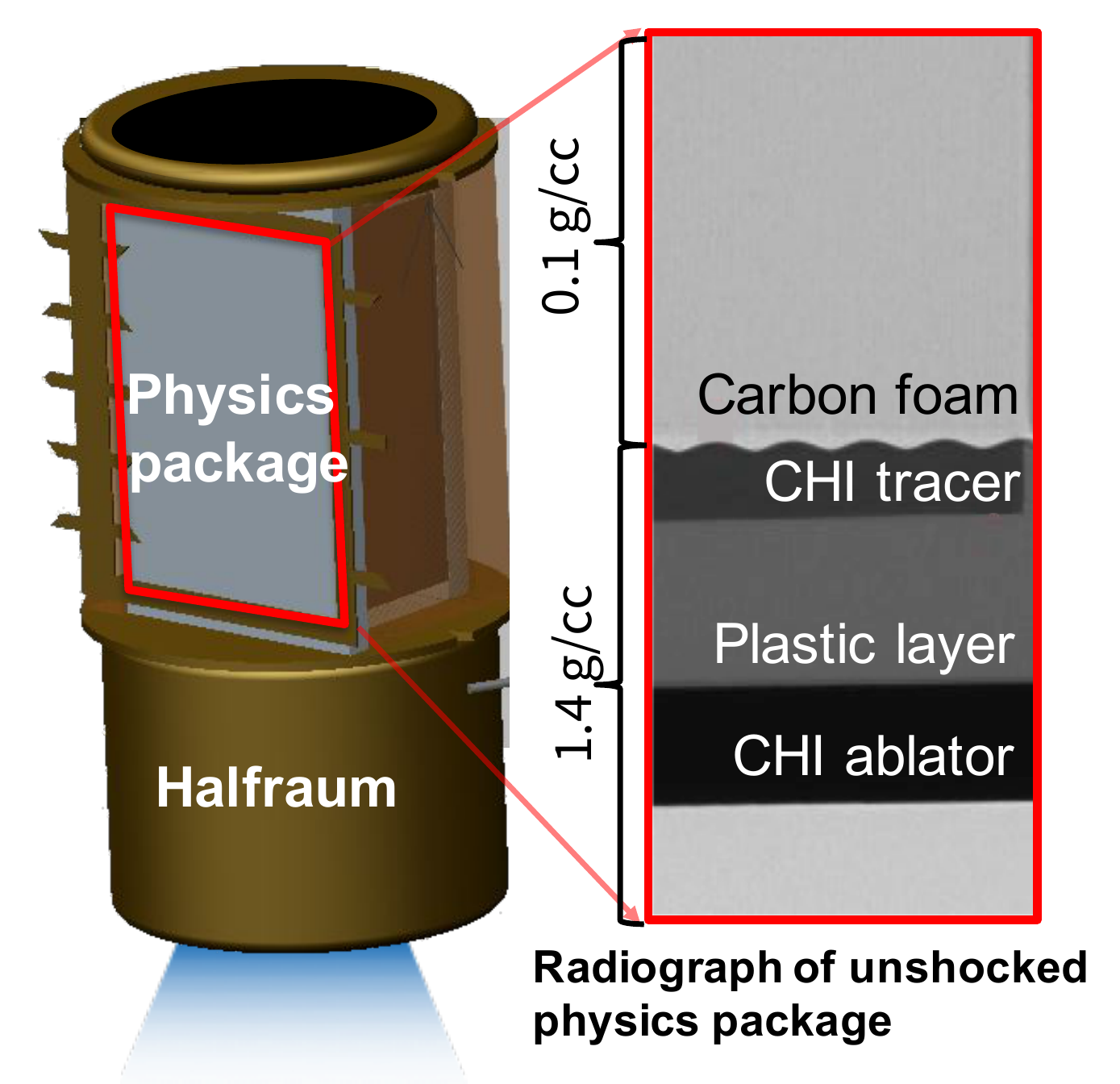}
 \caption{Left: schematic of the target package, Right: pre-shot radiograph of the physics package. Figure adapted from Nagel et al. \cite{Nagel:2017}}
 \label{fig:TargetLayout}
 \end{center}
 \end{figure}

\noindent \textbf{The Physics Package} The high density component, or ablator,  involves two density-matched (1.43 g/cm$^3$) plastics: polyamide-imide (PAI), C$_{22}$H$_{14}$N$_2$O$_3$, and an iodine-doped polystyrene (CHI), C$_{50}$H$_{47}$I$_3$.  Models and previous experiments suggest that the plastics behave hydrodynamically similar \cite{Nagel:2017}, but the CHI is relatively opaque to the imaging x-ray energy compared to the PAI.  Therefore, a 300 $\mu$m thick slice of CHI is placed at the center of the PAI to act as a tracer strip. This improves the contrast between the plastic and foam, and reduces multi-dimensional effects. The ablator compound is machined with a sinusoidal perturbation on the plastic-foam interface wavelength $\lambda$ = 200 $\mu$m and an initial sinusoidal amplitude ($a_{0}$) of  $a_{0}$ = 7.5 $\mu$m and 15 $\mu$m  respectively for two separate targets. Figure \ref{fig:TargetLayout} shows a sample pre-shot radiograph and drawing of the physics package. In the left image in Fig. \ref{fig:TargetLayout}, the lasers irradiate the halfraum generating an x-ray bath that ablates the plastic and drives a shock that propagates through the physics package. \cite{Nagel:2017} The shock accelerates the high density CHI ablator, PAI layer, and CHI tracer into the low density CRF foam, creating an RT unstable interface.

 According to the criteria for nonlinearity described in the above,  growth rates and dynamic structure begin to deviate from classical theory when $a_{0}/\lambda \sim$ 0.1 The initial conditions set on this platform obtain an initial  a$_{0}$/$\lambda$ ratio of 0.037 and 0.075 respectively and the experiment already begins in the weakly nonlinear regime. Radiographs of previous NIF experiments confirm that the perturbations grow into the deeply nonlinear regime by about halfway through the $\sim$ 60 ns experiment. \cite{Robey:2003, Nagel:2017} Figure \ref{fig:Reshock Data} (Right) shows that the interface has evolved from the sinusoidal initial condition ($\lambda$ =120  $\mu$m and a$_0$ = 6  $\mu$m), and has entered the nonlinear regime by 46 ns after the initial laser drive. 
 
 The spike tips display mushroom cap features characteristic of the late-time RT behavior that the Vortex experiment proposes to image. Increasing the wavelength from $\lambda$ =120 $\mu$m to $\lambda$ = 200 $\mu$m  increases the spacing between the spike tips and mitigate any possibility of the spike mergers.  To compensate for the large wavelength, the initial amplitudes in the Vortex campaign are increased to $a_{0}$ =  7.5 $\mu$m and 15 $\mu$m and compensates for the increased wavelength. The spike tip features in Fig. \ref{fig:Reshock Data} (Right) are notably indiscernible at the 28  $\mu$m spatial resolution, and a higher resolution x-ray diagnostic is required to diagnose the dynamics within that region.

\section{Radiography diagnostic} Previous NIF experiments, while well-characterized and having obtained the highest  Signal-to-Noise ratio (SNR) image of any laser-driven, single-mode HED instability to date, were not optimized for resolution. \cite{Nagel:2017} Figure \ref{fig:Reshock Data} shows data obtained from the  campaign conducted in 2017 \cite{Nagel:2017} using big-area-backlighter (BABL) scheme developed by Flippo et. al.  \cite{flippo2014development}  The spatial resolution of the system in Figure \ref{fig:Reshock Data} (right) is $\approx$ 28 $\mu$m, mainly set by the 25 $\mu$m pinhole diameter. \cite{flippo2014development}  The length scales observed in the vortex features along the spike tip shown in Figure \ref{fig:Reshock Data} (Right)  are less than 10 $\mu$m , and well below the 28 $\mu$m resolution.  Reducing the pinhole diameter to achieve a sufficient spatial resolution results in the fluence (photons per unit area) becoming prohibitively small. \cite{DoZonePlate}   The objective of this experiment is to observe fine scale structure in a well characterized RT unstable system that is optimized for a high-resolution diagnostic. Recently-commissioned diagnostics, such as the CBI,  provide greatly improved resolution ($\sim$3x better than current pinhole imaging, $\sim$7 $\mu$m vs $\sim$28 $\mu$m).

   \begin{figure}
 \center
  \includegraphics[width = 2.7in]{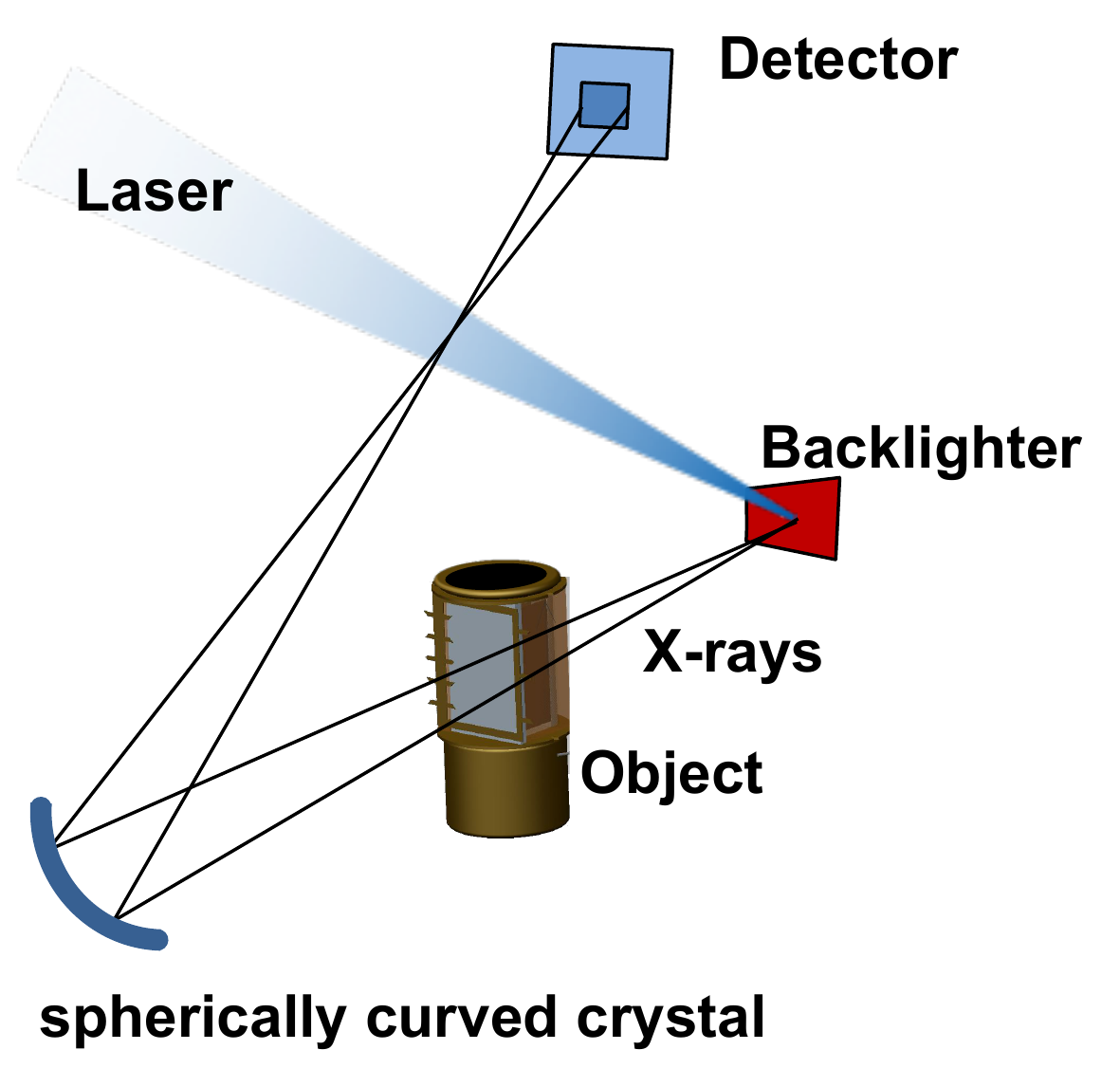}
  \caption{Schematic of a Crystal Backlighter Imager using a spherically bent crystal and a laser-created x-ray source on the Rowland circle. A NIF quad is incident on the Se Backlighter and emits 11.6 keV x-rays through the target that is then reflected off the crystal and onto an image plate. \cite{Hall:2019}}
  \label{fig:CBImodel}
\end{figure}

\noindent \textbf{Crystal X-Ray Imaging}
Multi-keV x-ray imaging coupled with a concave crystal x-ray mirror has become an invaluable diagnostic in the HED field.  Bent-crystal microscopes provide a high Signal-to-Noise ratio (SNR), high spatial resolution image over a cm-size field of view (FOV) by backlighting or imaging the target's self emission.  X-ray imaging with bent crystals relies on the Bragg diffraction of x-rays from crystal planes and only x-rays that satisfy the Bragg condition (Eqn.\ref{eqn:Bragg}) are reflected by the crystal lattice. The Bragg equation is 

\begin{equation}
E = \frac{mhc}{2dsin\theta},
\label{eqn:Bragg}
\end{equation}

where $E$ is the spectral line energy of the backlighter, $m$ is the reflection order, $h$ is Planck's constant, $c$ is the velocity of light in vacuum,  $\theta$ is the Bragg angle, and $d$ is the lattice spacing.

The CBI is a quasi-monochromatic, near-normal incidence, spherically bent crystal imager developed for the NIF. The CBI backlighter system  uses a 11.652 keV selenium He-$\alpha$ resonance line coupled to a silicon (8 6 2) crystal. \cite{Hall:2019,schollmeier} The backlighter target is a planar foil with a 10 $\mu$m layer of Se coated onto a 50 $\mu$m Al substrate that is driven by four NIF beams. Only the x-rays produced by the Se backlighter are reflected by Si crystal (Table 1).  This produces an image with a high SNR and large fluence of 11.6 keV X-rays, while obtaining a sub 10 $\mu$m resolution and magnification by a factor of ten. \cite{Hall:2019}  The length of the laser pulse used to generate the x-ray pulse  determines the temporal resolution of the experiment.

The CBI is presently used to radiograph ICF capsule implosions and has already demonstrated significant improvement over radiographs produced by pinhole imaging schemes.  The Vortex experiment is the first ns-time scale hydrodynamic experiment to be performed using the CBI diagnostic. While the present resolution limit for the CBI is 7 $\mu$m, future improvements to the diagnostic aim to reduce the spatial resolution to $\sim$ 1 $\mu$m.

\begin{table}
\center
\begin{tabular}{ll}
\hline
Crystal                                & Silicon 8 6 2                     \\
 \midrule
Atomic Line                               & Se He$_{\alpha}$   \\
 \midrule
X-ray energy                              & 11.6 keV   \\
 \midrule
Magnification                             & 10x  \\
 \midrule
Bragg angle                               & 86.6 deg \\
 \midrule
Spatial Resolution                        & 7 $\mu$m\\
 \midrule      
\end{tabular} 
 \label{table:diagnostic}
  \caption{Characteristics of the initial spherical CBI}

\end{table}

 \begin{figure*}
 \begin{center}
 \includegraphics[width = 6.5in]{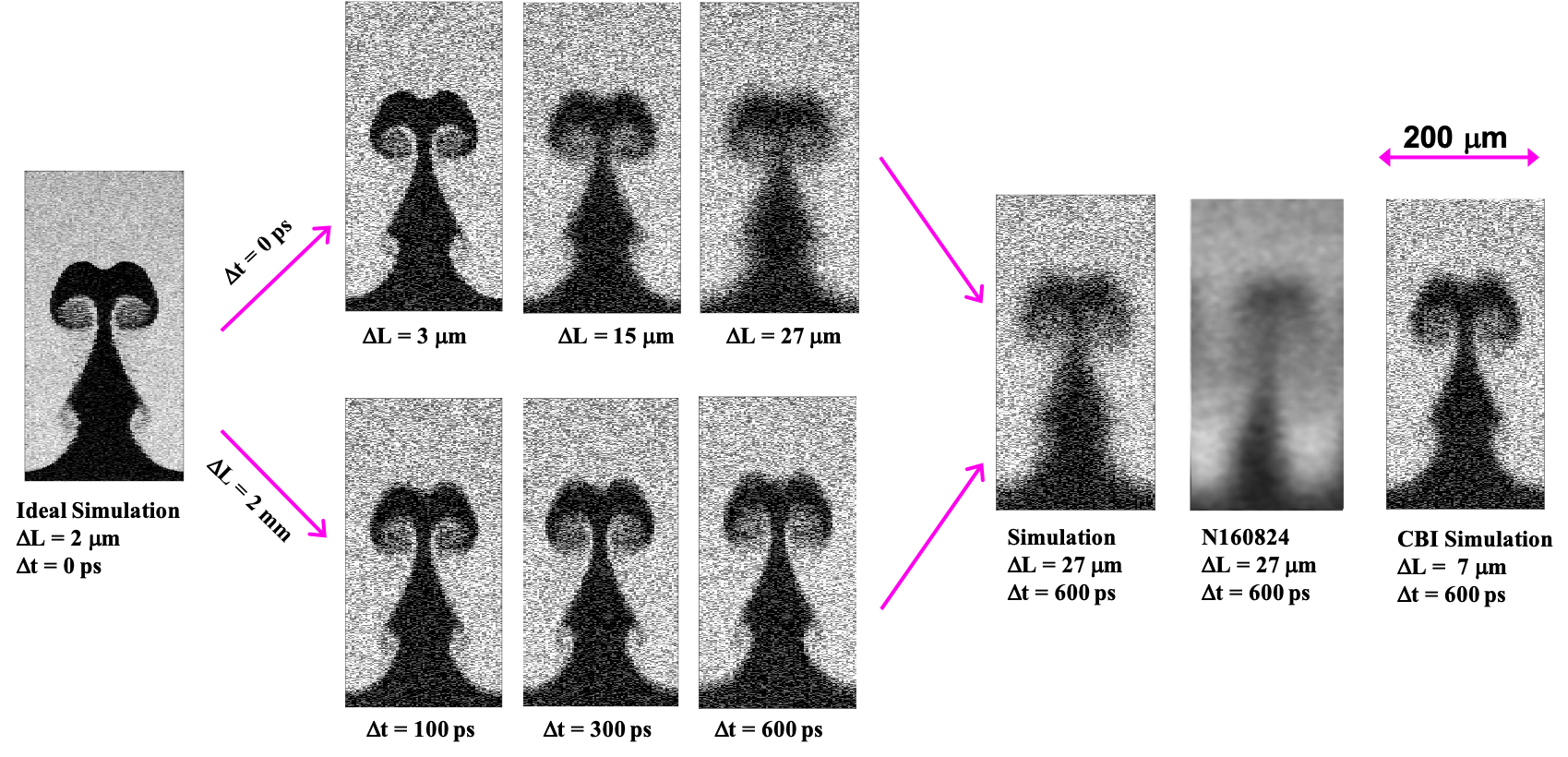}
 \caption{This figure shows a series of synthetic radiographs based on post-processed 2D simulations, assuming a monochromatic 9 keV x-ray energy, systematically degraded spatially and temporally until the synthetic radiograph is comparable to data taken with the BABL and what is achievable with the CBI. The ideal simulation has a spatial resolution ($\Delta$L) of 2 $\mu$m and temporal resolution ($\Delta$t) of 0 ps.  The figure then bifurcates where the top series of synthetic radiographs sets the temporal resolution as constant ($\Delta$t = 0) while the spatial resolution degrades, and the bottom series sets the spatial resolution constant ($\Delta$L = 2 $\mu$m) while the temporal resolution degrades.  The top row shows the effect of the various diagnostic capabilities on the experimental resolution.  The bottom row shows the contribution of motion blur and therefore the importance of moving to a shorter integration time ($\Delta t$). This converges into a synthetic radiograph emulating BABL conditions, example data from shot N160824-003, and finally a synthetic radiograph emulating CBI conditions. The CBI simulation incorporates the 11.6 keV x-ray energy in addition to the temporal and spatial resolutions listed above. The synthetic radiographs were post processed with blur ($\Delta$L = FWHM) and Poisson noise.
} 
 \label{fig:synthetic}
 \end{center}
 \end{figure*}

\section{Quantification of Synthetic Data}

Using the characterization of the CBI instrument described above we are able to produce ``synthetic radiographs" by post-processing 2D simulations with the appropriate resolution and noise properties of the instrument. We created the 2D simulations for this exercise using the radiation hydrodynamics code HYDRA. \cite{HYDRAcite, Marinak_2001} HYDRA is a multi-physics simulation code
from LLNL, that is used to simulate a variety of experiments carried out at NIF. HYDRA can simulate radiation transfer, atomic physics, hydrodynamics,
laser propagation, and a number of other physics effects. \cite{Marinak_2001}

The 2D hydrodynamic simulation performed represents a planar slice through the target, and we account for the third dimension--the axis parallel with the radiograph vector--by assuming that the 2D slice represents the material profile along this axis throughout the target. Non-planarity has a negligible impact on the measurement since the contrast in the x-ray radiograph is sensitive almost exclusively to the 300 $\mu$m CHI doped tracer strip. Small-scale details of the flow field, such as $\lambda_{\rm{LT}}$, vary at different planes within the 300 $\mu$m tracer strip. These details are not resolvable in the limit of the 2D approximation for radiographs with the 7 $\mu$m resolution of the CBI.  Therefore, further loss of image fidelity around the smallest-scale features of the image does not change our conclusions with the present platform. Future experimental designs with $\sim$ 1 $\mu$m capabilities can address this issue with a narrower tracer strip, a lower-energy x-ray probe to retain contrast, and additional metrics to reduce misalignment. We look forward to imaging diagnostics with sufficient resolution that we are faced with this problem.

The transmission in Fig. \ref{fig:synthetic} shows the x-ray transmission $I$ through the 2D simulation. The x-ray transmission calculation uses the map of material location (plastic and foam) at each point $(x,y)$, and the density at each point, $\rho(x,y)$. The transmission fraction $I(x,y, E)$ is calculated for each pixel of the synthetic radiograph according to:
\begin{align}
I(x,y,E) = I_0\prod_i e^{\left(-\rho(x,y)_i \times \sigma(E)_i \times l_i \right)},                                                                                            
\end{align}
where $\sigma$ is the mass attenuation coefficient, $l$ is the thickness of each material, and the subscript $i$ designates each of the materials along the x-ray line of sight. The length of foam ($l_{\rm{foam}}$) is 1900 $\mu$m everywhere that there is foam in the 2D map of material. Where the simulation indicates that there is plastic, it is assumed that there is 1600 $\mu$m of PAI ($l_{\rm{PAI}}$ = 1600 $\mu$m) and 300 $\mu$m of CHI ($l_{\rm{CHI}}$ = 300 $\mu$m) along the line of sight.  The mass attenuation coefficient, $\sigma$, depends only on the material and the x-ray energy, $E$.\cite{XCOM} All the synthetic radiographs in Figure \ref{fig:synthetic} calculate the transmission of the material based off the 9 keV x-ray energy used for the BABL, except the CBI Simulation which calculates the transmission using the 11.6 keV x-ray energy used for the CBI.

A qualitative comparison between the existing data can be made by artificially degrading the x-ray transmission image according to the diagnostic characteristics. The pinhole camera used to produce the data in Fig. \ref{fig:synthetic}  has been shown to have an image resolution of approximately 27 $\mu$m (full-width-half-max point spread function). \cite{Nagel:2017} The simulations produced in Fig. \ref{fig:synthetic} shows the result of performing a Gaussian blur of this magnitude that represents the diagnostic resolution, and imposing a level of Poisson noise to represent the signal noise at the scale of the image plate resolution element (25 $\mu$m), to the ideal image. Though the structure is not exact, this synthetic pinhole camera image is qualitatively similar to the experimental data; note, the data was collected at 9.0 keV. \cite{Nagel:2017} We can apply the same processes with the imaging parameters of the CBI, namely a FWHM resolution of 7 $\mu$m (the same magnitude of image-plate Poisson noise was applied). Although the fine details in the vortex regions at the spike tip and along the spike are still below the resolution limit, the improvement is significant.




\section{Conclusion}

 Many experiments have been conducted to study RT instabilities in HEDP over decades. Vorticity present along the RT spike tip not only causes the spike tip to broaden but can induce a transition to turbulence within the resulting vortex feature. A way to diagnose turbulence within this region is to for the flow to reach a sufficiently high $Re$, and to observe the presence of $\lambda_{\rm{LT}}$.  Facility and diagnostic advancements, such as the CBI, now allow us to reach regimes and image features that were previously unattainable. The Vortex experiment described in this paper adapts the platform of a well-characterized RT unstable system to the CBI to resolve spike tip morphology in a deeply non-linear RT system. Although the  resolution limit for the present campaign is $7 \mu m$, future improvements to the experimental platform can improve the diagnostic resolution while increasing the $\lambda_{\rm{LT}}$.  Simulations help inform the need to reduce motion blur ($\Delta$t) in order to achieve the desired resolution.

%




\section{Acknowledgements}
This work is funded by the U.S. Department of Energy NNSA Center of Excellence under cooperative agreement number DE-NA0003869. This work was performed under the auspices of the U.S. Department of Energy by Lawrence Livermore National Laboratory under contract DE-AC52-07NA27344. Lawrence Livermore National Security, LLC.  This document may contain research results that are experimental in nature, and neither the United States Government, any agency thereof, Lawrence Livermore National Security, LLC, nor any of their respective employees makes any warranty, express or implied, or assumes any legal liability or responsibility for the accuracy, completeness, or usefulness of any information, apparatus, product, or process disclosed, or represents that its use would not infringe privately owned rights. Reference to any specific commercial product, process, or service by trade name, trademark, manufacturer, or otherwise does not constitute or imply an endorsement or recommendation by the U.S. Government or Lawrence Livermore National Security, LLC. The views and opinions of authors expressed herein do not necessarily reflect those of the U.S. Government or Lawrence Livermore National Security, LLC and will not be used for advertising or product endorsement purposes.

 \bibliographystyle{ThreeAuthors.bst} 		
 \bibliography{VorteXDesignIFSA_v2}

\providecommand{\noopsort}[1]{}\providecommand{\singleletter}[1]{#1}%
\begin{thebibliography}{10}
\newcommand{\enquote}[1]{``#1''}
\expandafter\ifx\csname urlstyle\endcsname\relax
  \providecommand{\doi}[1]{doi:\discretionary{}{}{}#1}\else
  \providecommand{\doi}{doi:\discretionary{}{}{}\begingroup
  \urlstyle{rm}\Url}\fi

\bibitem{Hall:2019}
G.~N. Hall, C.~M. Krauland, M.~S. Schollmeier, et~al.
\newblock \enquote{The Crystal Backlighter Imager: A spherically bent crystal
  imager for radiography on the National Ignition Facility}.
\newblock \emph{Review of Scientific Instruments}, \textbf{90}(1), 013702
  (2019).
\newblock \doi{10.1063/1.5058700}.

\bibitem{DoZonePlate}
A.~Do, A.~M. Angulo, G.~N. Hall, et~al.
\newblock \enquote{X-ray imaging of Rayleigh Taylor instabilities using Fresnel
  zone plate at the National Ignition Facility}.
\newblock \emph{Review of Scientific Instruments}, \textbf{92}(5), 053511
  (2021).
\newblock \doi{10.1063/5.0043682}.

\bibitem{Remington:2006}
B.~A. Remington, R.~P. Drake, and D.~D. Ryutov.
\newblock \enquote{Experimental astrophysics with high power lasers and $Z$
  pinches}.
\newblock \emph{Rev. Mod. Phys.}, \textbf{78}, 755 (2006).
\newblock \doi{10.1103/RevModPhys.78.755}.

\bibitem{Robey:2003}
H.~F. Robey, Y.~Zhou, A.~C. Buckingham, et~al.
\newblock \enquote{The time scale for the transition to turbulence in a high
  Reynolds number, accelerated flow}.
\newblock \emph{Physics of Plasmas}, \textbf{10}(3), 614 (2003).
\newblock \doi{10.1063/1.1534584}.

\bibitem{RPD}
R.~P. Drake, E.~C. Harding, and C.~C. Kuranz.
\newblock \enquote{Approaches to turbulence in high-energy-density
  experiments}.
\newblock \emph{Physica Scripta}, \textbf{T132}, 014022 (2008).
\newblock \doi{10.1088/0031-8949/2008/t132/014022}.

\bibitem{Nagel:2017}
S.~R. Nagel, K.~Raman, C.~M. Huntington, et~al.
\newblock \enquote{A platform for studying the rayleigh-taylor and
  richtmyer-meshkov instabilities in a planar geometry at high energy density
  at the national ignition facility}.
\newblock \emph{Physics of Plasmas}, \textbf{913}, 103 (2019).
\newblock ISSN 0168-9002.
\newblock \doi{10.1016/j.nima.2018.10.119}.

\bibitem{Smalyuk:2017}
V.~Smalyuk, H.~Robey, D.~Casey, et~al.
\newblock \enquote{Mix and hydrodynamic instabilities on NIF}.
\newblock \emph{Journal of Instrumentation}, \textbf{12}, C06001 (2017).
\newblock \doi{10.1088/1748-0221/12/06/C06001}.

\bibitem{Hurricane_2016}
O.~A.~H. and.
\newblock \enquote{Overview of Progress and Future Prospects in Indirect Drive
  Implosions on the National Ignition Facility}.
\newblock \emph{Journal of Physics: Conference Series}, \textbf{717}, 012005
  (2016).
\newblock \doi{10.1088/1742-6596/717/1/012005}.

\bibitem{Clark_2019}
D.~Clark, C.~Weber, J.~Milovich, et~al.
\newblock \enquote{Three-dimensional modeling and hydrodynamic scaling of
  National Ignition Facility implosions}.
\newblock \emph{Physics of Plasmas}, \textbf{26}, 050601 (2019).
\newblock \doi{10.1063/1.5091449}.

\bibitem{darlington2001study}
R.~M. Darlington, T.~L. McAbee, and G.~Rodrigue.
\newblock \enquote{A study of ALE simulations of Rayleigh--Taylor instability}.
\newblock \emph{Computer Physics Communications}, \textbf{135}(1), 58 (2001).

\bibitem{miles2004bubble}
A.~Miles.
\newblock \enquote{Bubble merger model for the nonlinear Rayleigh--Taylor
  instability driven by a strong blast wave}.
\newblock \emph{Physics of plasmas}, \textbf{11}(11), 5140 (2004).

\bibitem{Zhou:2017}
Y.~Zhou.
\newblock \enquote{Rayleigh–Taylor and Richtmyer–Meshkov instability
  induced flow, turbulence, and mixing. II}.
\newblock \emph{Physics Reports}, \textbf{723-725}, 1  (2017).
\newblock ISSN 0370-1573.
\newblock \doi{https://doi.org/10.1016/j.physrep.2017.07.008}.
\newblock Rayleigh–Taylor and Richtmyer–Meshkov instability induced flow,
  turbulence, and mixing. II.

\bibitem{Rayleigh}
L.~Rayleigh.
\newblock \enquote{Scientific Papers II, 200}.
\newblock \emph{Cambridge, England} (1900).

\bibitem{taylor1950instability}
G.~I. Taylor.
\newblock \enquote{The instability of liquid surfaces when accelerated in a
  direction perpendicular to their planes. I}.
\newblock \emph{Proceedings of the Royal Society of London. Series A.
  Mathematical and Physical Sciences}, \textbf{201}(1065), 192 (1950).

\bibitem{jacobs1996experimental}
J.~W. Jacobs and J.~Sheeley.
\newblock \enquote{Experimental study of incompressible Richtmyer--Meshkov
  instability}.
\newblock \emph{Physics of Fluids}, \textbf{8}(2), 405 (1996).

\bibitem{roberts2016effects}
M.~Roberts and J.~W. Jacobs.
\newblock \enquote{The effects of forced small-wavelength, finite-bandwidth
  initial perturbations and miscibility on the turbulent Rayleigh--Taylor
  instability}.
\newblock \emph{Journal of Fluid Mechanics}, \textbf{787}, 50 (2016).

\bibitem{zhou2019turbulent}
Y.~Zhou, T.~T. Clark, D.~S. Clark, et~al.
\newblock \enquote{Turbulent mixing and transition criteria of flows induced by
  hydrodynamic instabilities}.
\newblock \emph{Physics of Plasmas}, \textbf{26}(8), 080901 (2019).

\bibitem{robey2003onset}
H.~Robey, Y.~Zhou, A.~Buckingham, et~al.
\newblock \enquote{The onset of turbulence in high reynolds number, accelerated
  flows. part ii. experiment}.
\newblock \emph{Physics of Plasmas}, \textbf{10} (2003).

\bibitem{dimotakis2000mixing}
P.~E. Dimotakis.
\newblock \enquote{The mixing transition in turbulent flows}.
\newblock \emph{Journal of Fluid Mechanics}, \textbf{409}, 69 (2000).

\bibitem{dimonte2006k}
G.~Dimonte and R.~Tipton.
\newblock \enquote{K-L turbulence model for the self-similar growth of the
  Rayleigh-Taylor and Richtmyer-Meshkov instabilities}.
\newblock \emph{Physics of Fluids}, \textbf{18}(8), 085101 (2006).

\bibitem{zhou2003progress}
Y.~Zhou, B.~Remington, H.~Robey, et~al.
\newblock \enquote{Progress in understanding turbulent mixing induced by
  Rayleigh--Taylor and Richtmyer--Meshkov instabilities}.
\newblock \emph{Physics of Plasmas}, \textbf{10}(5), 1883 (2003).

\bibitem{zhou2007unification}
Y.~Zhou.
\newblock \enquote{Unification and extension of the similarity scaling criteria
  and mixing transition for studying astrophysics using high energy density
  laboratory experiments or numerical simulations}.
\newblock \emph{Physics of Plasmas}, \textbf{14}(8), 082701 (2007).

\bibitem{flippo2014development}
K.~Flippo, J.~Kline, F.~Doss, et~al.
\newblock \enquote{Development of a Big Area BackLighter for high energy
  density experiments}.
\newblock \emph{Review of Scientific Instruments}, \textbf{85}(9), 093501
  (2014).

\bibitem{schollmeier}
M.~Schollmeier and G.~Loisel.
\newblock \enquote{Systematic search for spherical crystal X-ray microscopes
  matching 1–25 keV spectral line sources}.
\newblock \emph{The Review of scientific instruments}, \textbf{87}, 123511
  (2016).
\newblock \doi{10.1063/1.4972248}.

\bibitem{HYDRAcite}
S.~H. Langer, I.~Karlin, and M.~M. Marinak.
\newblock \enquote{Performance Characteristics of HYDRA --A Multi-physics
  Simulation Code from LLNL}.
\newblock In M.~Dayd{\'e}, O.~Marques, and K.~Nakajima, editors, \emph{High
  Performance Computing for Computational Science -- VECPAR 2014}, pages
  173--181. Springer International Publishing, Cham (2015).
\newblock ISBN 978-3-319-17353-5.

\bibitem{Marinak_2001}
M.~M. Marinak, G.~D. Kerbel, N.~A. Gentile, et~al.
\newblock \enquote{Three-dimensional HYDRA simulations of National Ignition
  Facility targets}.
\newblock \emph{Physics of Plasmas}, \textbf{8}(5), 2275 (2001).
\newblock \doi{10.1063/1.1356740}.

\bibitem{XCOM}
B.~Henke, E.~Gullikson, and J.~Davis.
\newblock \enquote{X-ray interactions: photoabsorption, scattering,
  transmission, and reflection at E=50-30000 eV, Z=1-92}.
\newblock \emph{tomic Data and Nuclear Data Tables}, \textbf{54}(2), 181
  (1993).
\newblock \doi{10.1063/5.0043682}.

\end{thebibliography}

\end{document}